
\magnification=\magstep1
\tolerance=2000

\font\large=cmbx10 at 12 pt
\newcount\equationno      \equationno=0
\newtoks\chapterno \xdef\chapterno{}
\def\eqn{\eqno\eqname}
\def\eqnx{\eqname} 
\def\eqname#1{\global \advance \equationno by 1 \relax
\xdef#1{{\noexpand{\rm}(\chapterno\number\equationno)}}#1}

\def\la{\mathrel{\mathchoice {\vcenter{\offinterlineskip\halign{\hfil
$\displaystyle##$\hfil\cr<\cr\sim\cr}}}
{\vcenter{\offinterlineskip\halign{\hfil$\textstyle##$\hfil\cr<\cr\sim\cr}}}
{\vcenter{\offinterlineskip\halign{\hfil$\scriptstyle##$\hfil\cr<\cr\sim\cr}}}
{\vcenter{\offinterlineskip\halign{\hfil$\scriptscriptstyle##$\hfil\cr<\cr\sim
\cr}}}}}


\centerline{\large Response of finite-time particle detectors in}
\medskip
\centerline{\large non-inertial frames and curved spacetime}
\vskip 1truein
\centerline{\bf L. Sriramkumar$^1$ and T. Padmanabhan$^2$}
\centerline{IUCAA, Post Bag 4, Ganeshkhind,}
\centerline{Pune 411007, INDIA.}
\vskip 1truein
\centerline{\bf Abstract}
\bigskip
\noindent The response of the Unruh-DeWitt type monopole detectors which were
coupled to the quantum field only for a finite proper time interval is studied
for inertial and accelerated trajectories, in the Minkowski vacuum in (3+1)
dimensions. Such a detector will respond even while on an inertial trajctory
due to the transient effects. Further the response will also depend on the
manner in which the detector is switched on and off. We consider the response
in the case of smooth as well as abrupt switching of the detector. The former
case is achieved with the aid of smooth window functions whose width, $T$,
determines the effective time scale for which the detector is coupled to the
field. We obtain a general formula for the response of the detector when a
window function is specified, and work out the response in detail for the case
of gaussian and exponential window functions. A detailed discussion of both
$T \rightarrow 0$ and $T \rightarrow \infty$ limits are given and several
subtlities in the limiting procedure are clarified. The analysis is extended
for detector responses in Schwarzschild and de-Sitter spacetimes in (1+1)
dimensions.

\vskip 1truein
\centerline{{\bf IUCAA -- 23/94, August, 1994} : Submitted for publication}
\vfill
$^1$ sriram@iucaa.ernet.in

$^2$ paddy@iucaa.ernet.in
\vfill\eject

\beginsection{\bf 1. Introduction}

In studying the quantum field theory in Minkowski spacetime, we identify the
coefficients of the positive frequency component of the field modes to be the
annihilation operators and define the state that gets annihilated by these
operators to be the vacuum state for the field. This theory being invariant
under the Poincare group, the vacuum state defined by this prescription will
be the same for all inertial observers. But the vacuum defined by this
procedure is not invariant under a general coordinate transformation in flat
spacetime. It is well known, for example that quantisation in Minkowski
coordinates and Rindler coordinates are not equivalent$^{[1]}$. This problem
araises again while studying quantum fields in a given curved spacetime:
the vacuum state and the particle concept are not invariant under general
coordinate transformations while the classical field theory is. Concepts like
`vacuum', `particles' etc.,  defined through conventional quantum field
theoretic methods do not seem to posess any universal significance but rather
have an observer dependant quality about them.
\smallskip
The concept of particle detectors was introduced into this subject$^{[2, 3]}$,
with the goal of improving our understanding of the concept of a particle in
an arbitrary curved spacetime. The general philosophy was that:``Particles
are what the particle detectors detect"$^{[4]}$. With this motivation,
detectors coupled to quantum fields were designed and their responses were
studied. Though this has been done extensively in literature, only a limited
amount of insight into field theory in curved spacetime seems to have been
acquired in the process.
\smallskip
The response of these detectors is usually studied for their entire history,
{\it viz} from the infinite past to the infinite future in the detector's
proper time. But in any realistic situation the detectors can be kept
switched on only for a finite period of time and in this context the study
of the response of the detector during a finite interval in proper time
gains importance. There has been a couple of attempts in literature in the
recent past, when the finite time detector response was calculated$^{[5,
6]}$. The authors of these papers, however, have encountered certain divergent
results which are difficult to interpret physically. The authors in $[5]$
resort to a complicated `renormalisation' procedure to remove the divergences
while in $[6]$ an attempt is made to eliminate the divergences using a
smooth window function.
\smallskip
We reanalyse this question in the present paper. We begin by noting that a
detector which is ``kept on" only for a finite interval $T$ will be affected
by the transients related to the process of switching. This has the consequence
that, even an inertial detector in the Minkowski vacuum will register a
response
for finite $T$. This effect, as we shall see, needs to be clearly identified
before one studies the response in an accelerated trajectory for finite $T$.
Further, we expect the response to vanish when $T \rightarrow 0$ for {\it any}
realistic detector on {\it any} trajectory. This is simply a physical
requirement arising from the demand that ``a detector which was never switched
on should not detect anything". While this demand sounds reasonable, its
mathematical implementation turns out to be fairly subtle. We will see that
spurious results can arise if one does not implement the limiting procedure
with care. When they are done properly no divergences appear and the results
turn out to be physically reasonable.
\smallskip
The response of a detector depends, in general, on the following three
elements: (1) the state of the quantum field, (2) the trajectory of the
detector and (3) the nature of coupling that exists between the field and the
detector. In this paper, we assume that the coupling between the detector and
the field is of the linear monopole type$^{[2, 3]}$. We consider inertial and
accelerated trajectories with the field being in the Minkowski vacuum.
\smallskip
This paper is organised as follows. In section {\bf 2} we review the monopole
detector theory and comment on certain limiting procedures. In section {\bf 3}
we study the response of the detector which is operational only for a finite
interval of time; the case of a smooth window function as well as that of
abrupt switching on is considered. Section {\bf 4} dicusses the extension
of the finite time detector response theory analysed in the earlier sections
to Schwarzschild and de-Sitter spacetimes. Section ${\bf 5}$ discusses
possible conclusions from our analysis.
\bigskip
\beginsection{\bf 2. Response of Unruh-DeWitt detector - revisited}

In this section we study the case of a massless, minimally coupled scalar
field in (3+1) or (1+1) dimensions with the field being assumed to be in the
Minkowski vacuum. The detector-field interaction is described by the
interaction lagrangian of the form $c\;m(\tau)\;\Phi[x(\tau)]$, where $c$ is
a small coupling constant and $m(\tau)$ is the detector's monopole operator.
For a general trajectory, the detector will not remain in its ground state
$E_0$ but will undergo a transition to an excited state $E$ due to its
interaction with the scalar field. The amplitude for transition in the first
order of perturbation theory is

$${\cal A}\; =\; ic\; <E, \Psi\vert \int_{-\infty}^\infty d\tau\; m(\tau)\;
\Phi[x(\tau)]\; \vert 0_M, E_0>\eqn\qq.$$
Using the equation for the time evolution of $m(\tau)$,

$$m(\tau)\; =\;e^{i H_0 \tau}\; m(0)\; e^{-i H_0 \tau}\eqn\qq$$
where $H_0 \vert E>\; =\; E \vert E>$, the above transition amplitude
factorises to

$${\cal A}\; =\; {\cal M} \int_{-\infty}^\infty d\tau\; e^{i (E-E_0)\tau}
<\Psi\vert\Phi(x)\vert 0_M>\eqn\qamp$$
where $${\cal M}=i c <E \vert m(0) \vert E_0>\eqn\qq$$
with $\vert 0_M>$ denoting the Minkowski vacuum and $x(\tau)$ is the location
of the detector at proper time $\tau$.
\smallskip
If $\Phi$ is expanded in terms of the standard Minkowski plane wave modes,
it is clear from equation \qamp\ that the non-zero contribution to the
amplitude arises only from the state $\vert \Psi>\; =\; \vert 1_k>$. For
the case of an inertial trajectory in (1+1) dimensions, with

$$x(\tau)\; =\; x_0\; +\; vt\; =\; x_0\; +\; v \gamma \tau \eqn\qintraj$$
where $\gamma= (1-v^2)^{1/2}$, $x_0$ and  $v$ are constants, $\vert v
\vert< 1$, the amplitude \qamp\ turns out to be

$${\cal A}_{ine}\; =\; {\cal M}\; {e^{-i k x_0} \over {\sqrt{4 \pi \omega}}}
\int_{-\infty}^\infty d\tau\; e^{i(E-E_0) \tau}\; e^{i\gamma \tau (\omega -
k v)}\eqn\qq$$
with $\omega = \vert k \vert$. The integral gives a Dirac delta function and
we get

$${\cal A}_{ine}\; =\; {\cal M}\; {e^{-i k x_0} \over {\sqrt{4 \pi \omega}}}
\; \delta(a)\; = \; 0\eqn\qq$$
where $a\; \equiv\; (E\; -\; E_0\; +\; \gamma(\omega - k v))$. The last
equality follows from noting that since, $k\;v\; \leq\; \vert k \vert\;
\vert v \vert\; <\; \omega$ and $E > E_0$, the argument of the $\delta$-
function is always greater than zero. The transition in the detector being
essentially forbidden on the grounds of energy conservation.
\smallskip
The following points should be stressed regarding the above  - apparently
trivial - calculation: (i) the amplitude is being calculated for the system
to make a transition from the state $\vert E_0 >$ in the {\it infinite past},
to the state $\vert E >$ in the {\it infinite future}. To do so we need to
know the trajectory $x^i(\tau)$ for all $\tau$; {\it i. e} for $-\infty
< \tau < \infty$. No realistic detector can be kept switched on forever.
Suppose the detector was kept switched on only during  the time interval
$-T \le \tau \le T$; {\it then the amplitude will be non-zero}:

$${\cal A}_{ine}(T)\; =\; {\cal M}\; {e^{-i k x_0} \over {\sqrt{4 \pi
\omega}}} \int_{-T}^T d\tau\; e^{i(E-E_0) \tau}\; e^{i \gamma \tau (\omega
- k v)}\eqn\qq$$

$$ \; =\; {\cal M}\; {e^{-i k x_0} \over {\sqrt{4 \pi \omega}}}\; \left
\lbrace{2 \; \sin(a\; T) \over {(a\; T)}}\right\rbrace\eqn\qq.$$
The probability for transtition with a fixed $\omega$ will be

$${\cal P}_{ine, \omega}(T)\; =\; {\vert {\cal A}_{ine}(T) \vert}^2\; =\;
\left\lbrace{{\vert {\cal M} \vert}^2 \over {\pi \omega}}\right\rbrace\;
{\left\lbrace{\sin(a\; T) \over a}\right\rbrace}^2\eqn\qprofit$$
which is finite for all finite $T$. For small $T$, ${\cal P}_{ine, \omega}
\propto T^2$ and hence vanishes for $T\rightarrow 0$; for large $T$, we use
the relations

$$\lim_{T \rightarrow \infty} {\left\lbrace{\sin(a\;T) \over {\pi \; a}}
\right\rbrace}^2\; = \; \lim_{T \rightarrow \infty} \left\lbrace\biggl
(\lim_{T \rightarrow \infty} {\sin(a\; T) \over {\pi a}}\biggl) {\sin(a\;
T) \over {\pi a}}\right\rbrace $$
$$\; = \; \lim_{T \rightarrow \infty} \left\lbrace{\delta(a) \sin (a \;
T) \over {\pi a}}\right\rbrace\; =  \; \lim_{T \rightarrow \infty} \left
\lbrace {T \over \pi} \delta(a)\right\rbrace\eqn\qq.$$
In other words

$$\lim_{T \rightarrow \infty} \left\lbrace{{\cal P}_{ine, \omega}(T) \over
T}\right\rbrace\; =\; \left\lbrace{{\vert {\cal M} \vert}^2 \over {\omega}}
\right\rbrace \delta(a)\eqn\qq.$$
Clearly the rate of transitions ${\cal R}_{ine, \omega}(T)\; =\; {\cal P}_
{ine, \omega}(T) / T$ has the following behaviour: ${\cal R}_{ine, \omega}
\propto T$ for small $T$ and ${\cal R}_{ine, \omega} \propto \delta(a)$ for
large $T$. Hence ${\cal R}_{ine, \omega}$ vanishes in both the limits.
\smallskip
The above analysis should teach us three lessons: Firstly, even an inertial
detector will ``detect particles" if it is switched on and off. This is
merely a manifestation of the energy-time uncertainty principle; a detection
process lasting for a time $2T$ cannot measure energy differences with an
accuracy greater than $(2T)^{-1}$. So for $(a\; 2T)\; \la\; 1$, the rate
${\cal R}$ will be significantly non-zero. Secondly, the rate ${\cal R}$
is a more reliable quantity to compute than ${\cal P}$, especially if one
is considering the $T \rightarrow \infty$ limit. In particular, ${\cal P}$
is infinite if we take $T \rightarrow \infty$ limit naively in \qprofit .
Thirdly, if we want to study the response of accelerated detectors which are
switched on only for a finite time, we should subtract out the finite result
which is already present in the inertial case. The limits also need to be
handled with care to obtain sensible results. We shall say more about it
later on.
\smallskip
For the case of an uniformly accelerated trajectory in (1+1) dimensions,
the tranformations from the Minkowski to the accelerated frame are

$$x\; =\; g^{-1}\xi\; \cosh(g\tau);\qquad t\; =\; g^{-1}\xi\; \sinh (g \tau);
\eqn\qrincoords$$
where $\tau$ is the proper time of the accelerated observer at $\xi$. In what
follows we shall set $\xi=1$ without any loss of generality. The transition
amplitude for the accelerated trajectory of the detector turns out to be

$${\cal A}_{acc}\; =\; {{\cal M} \over {\sqrt {4 \pi \omega}}} \int_{-\infty}
^{\infty} d\tau\; e^{i(E-E_0)\tau}\; e^{-ikg^{-1} \cosh{g\tau} + i\omega
g^{-1} \sinh{g\tau}}\eqn\qq.$$
The above integral can be written down in terms of Gamma functions:

$${\cal A}_{acc}\; =\; {{\cal M} \over {\sqrt {4 \pi \omega}}}\; g^{-1}\;
e^{{- \pi\Omega} \over 2g} {(\omega g^{-1})}^{i\Omega g^{-1}}\; \Gamma(-i
\Omega g^{-1})\eqn\qq$$
where $\Omega = E - E_0$. This is clearly non-zero. The probalility for
transition ${\cal P}_{\omega}\; =\; {\vert {\cal A} \vert}^2$ with a fixed
$\omega$ will be

$${\cal P}_{acc, \omega}\; =\; {\vert {\cal A}_{acc} \vert}^2 = {{{\vert
{\cal M} \vert}^2} \over {4 \pi \omega}} \; {1 \over {g^2}} \; e^{- \pi
\Omega g^{-1}} {\vert \Gamma(-i \Omega g^{-1}) \vert}^2\; =\; {{{\vert {\cal
M} \vert}^2} \over {4 \pi \omega}}\; \left\lbrace{2 \pi \over {\Omega g}}\;
{1 \over {(e^{2 \pi \Omega g^{-1}} - 1)}}\right\rbrace\eqn\qthr$$
which has a Planckian form in $\Omega$ with temperature $\beta^{-1}\; =\;
({g \over {2 \pi}})$.
\smallskip
The finite proper time integral for the transition amplitude  for the
accelerated trajectory, obtained after substituting for $x$ and $t$ from
\qrincoords\ is

$${\cal A}_{acc}(T)\; =\; {{\cal M} \over {\sqrt{4 \pi \omega}}}\;
J\eqn\qq$$
where

$$ J\; =\; \int_{-T}^T d\tau\; e^{i \Omega \tau}\; e^{-i\omega g^{-1}
(\cosh{g\tau}-\sinh{g\tau})}\eqn\qq.$$
This integral for $J$ can be rewritten as

$$J\; =\; \int_{-\infty}^\infty d\tau\; e^{i\Omega \tau}\; e^{-i\omega
g^{-1} e^{-g\tau}}\; - \; \int_{-\infty}^T d\tau\; e^{i\Omega \tau}\;
e^{-i \omega g^{-1} e^{-g\tau}}\; -\; \int_T^\infty d\tau\; e^{i\Omega
\tau}\; e^{-i \omega g^{-1} e^{-g\tau}}\eqn\qq.$$
Each of the above integrals can be expressed in closed form as

$$J\; =\; \left\lbrace g^{-1}\; e^{-\pi \Omega \over 2g}\; {(\omega g^{-1})}
^{-i\Omega g^{-1}}\right\rbrace \; \biggl\lbrace \Gamma(-i\Omega g^{-1})
\; - \; \gamma(-i\Omega g^{-1}\; ,\; i\omega g^{-1} s^{-1}) \;$$
$$- \; \Gamma(-i\Omega g^{-1}\; ,\; i\omega g^{-1}s)\biggl\rbrace\eqn\qq$$
where $s\; =\; e^{gT}$, $\Gamma(m)$ is the complete gamma function and
$\Gamma(m,n)$ and $\gamma(m,n)$ are the incomplete gamma functions$^{[7]}$.
Consider now the limit $T \rightarrow 0$, ({\it i.e} when the detector is
not switched on at all). In this limit, $s \rightarrow 1$ and the two
incomplete gamma functions add upto the complete gamma function thereby
giving $J\; =\; 0$.  In the other limit, $T \rightarrow \infty$, $s
\rightarrow \infty$, $s^{-1} \rightarrow 0$ we get

$$J\; =\; g^{-1}\; e^-{\pi \Omega \over 2g}\; {(kg^{-1})}^{i\Omega g^{-1}}\;
\Gamma(-i \Omega g^{-1})\eqn\qq.$$
Evaluating ${\vert J \vert}^2$ we obtain the thermal spectrum in \qthr\ .
Thus we obtain reasonable results for both the limits $T \rightarrow 0$
as well as $T \rightarrow \infty$.
\smallskip
There is another feature that needs emphasis as regards both \qthr\ and
\qprofit\ . These are probabilities for transition to fixed final states
$\vert 1_{k}>$ charecterised by a given momentum k. Normally one would like
to integrate over all k so as to find the net probability for the detector
to have made a transition from $\vert E >$ to $\vert E_{0}>$. This will
lead to an integral

$$I_{ine}\; =\; \int_{0}^{\infty}{d\omega \over \omega}\; {\left\lbrace{
\sin((\Omega + \omega)T) \over {(\Omega + \omega)}}\right\rbrace}^2\eqn
\qq$$
in the case of \qprofit\ and to an integral

$$I_{acc}\; =\; \int_{0}^{\infty}{d\omega \over \omega}\; =\; \lim_{T_{1}
\rightarrow {\infty}}\lim_{T_{2} \rightarrow 0}\left\lbrace ln\left( {T_{1}
\over {T_{2}}}\right) \right\rbrace\eqn\qq$$
in the case of \qthr\ . Both these integrals are formally divergent.
However, consider the limit

$$\lim_{T \rightarrow \infty} \; \left\lbrace{I_{inertial} \over T}\right
\rbrace = \; \int_{0}^{\infty} {d\omega \over \omega} \; \left\lbrace
{\lim_{T \rightarrow \infty}} \; \left( {1 \over T} \; {\left( {\sin((\Omega
+ \omega)T) \over {(\Omega + \omega)}}\right) }^2 \right) \right\rbrace$$
$$= \; \int_{0}^{\infty} {d\omega \over \omega} \; \left\lbrace {1 \over \pi}
\delta(\Omega + \omega)\right\rbrace\eqn\qq.$$
If $\Omega > 0$, $\omega > 0$ the integrand identically vanishes and we may
take  this integral to be zero, thereby recovering the earlier result.
Exactly the same phenomenon takes place in the case of an accelerated
detector.
\smallskip
The probability of transition to all possible $E$ and $\vert \Psi>$ from
$E_0$ and $\vert 0_M>$, can be expressed in a more formal and concise manner
as:

$${\cal P}\; =\; \sum_{E,\; \vert\Psi>}{\vert {\cal A} \vert}^2\; =\;
\sum_E {\vert{\cal M}\vert}^2\; {\cal F}(\Omega)\eqn\qprob$$
with

$${\cal F}(\Omega)\; =\; \int_{-\infty}^\infty d\tau\; \int_{-\infty}^\infty
d\tau'\; e^{-i\Omega(\tau-\tau')}\; G^{+}(x(\tau), x(\tau'))\eqn\qdetres .$$
The detector response function ${\cal F}(\Omega)$, is independent of the
details of the detector and is determined completely by the positive
frequency Wightman function $G^{+}(x(\tau), x(\tau'))$ defined to be

$$G^{+}(x(\tau), x(\tau')) = <0_M\vert \Phi(x) \Phi(x')\vert 0_M> \eqn\qq.$$
The detector response function ${\cal F}(\Omega)$ represents the bath of
particles it experiences due to its motion. The remaining factor in \qprob\
represents the selectivity of the detector to this bath and depends on the
internal structure of the detector.
\smallskip
For trajectories in Minkowski space, which are integral curves of time-like
Killing vector fields (for e.g the inertial and the accelerated trajectories)
the Wightman function is invariant under time translations in the reference
frame of the detector$^{[8]}$. Hence

$$G^{+}(x(\tau), x(\tau'))\; =\; G^{+}(\tau-\tau')\; =\; G^{+}(\Delta (\tau))
\eqn\qq$$
And the double integration in \qdetres\ reduces to a Fourier transform of
the two point function multiplied by an infinite time interval. This
divergence is usually handled by interpreting the Fourier transform of the
two point function to be the transition probability per unit time,
{\it i.e} the rate is given by

$${\cal R}(\Omega)\; =\; \sum_E {\vert{\cal M}\vert}^2 \int_{-\infty}^
\infty d\Delta \tau\; e^{-i(E-E_0)\Delta \tau}\; G^{+}(\Delta \tau)\eqn
\qprobrate .$$
The Wightman function in the (3+1) dimensions for our field is $^{[9]}$
$$G^{+}(x,x')\; =\; -{1 \over 4 \pi^2 ({( t - t'- i\epsilon)}^2 -{\vert
{\bf x} -{\bf x'} \vert}^2)}\eqn\qmgfn$$
which, for the case of an inertial trajectory given by \qintraj\ ,reduces
to
$$G_{ine}^{+}(\Delta \tau)\; =\; -{1 \over 4 \pi^2 (\Delta \tau - i\epsilon)
^2} \eqn\qgfni .$$
(We have absorbed a positive factor $\gamma$ into $\epsilon$). Since $E>
E_0$, the integral \qprobrate\ can be performed by closing the contour in
an infinite semi-circle in the lower-half plane. But the pole of the two
point function \qgfni\ being at $\Delta \tau = i \epsilon$, it does not
contribute to the integral and the detector response is zero; in other
words the inertial detector does not see any particles in the Minkowski
vacuum.
\smallskip
For the case of an accelerated trajectory given by \qrincoords\ the Wightman
function is

$$G_{acc}^{+}(\Delta \tau)\; =\; -\left\lbrace {16 \pi^2 g^{-2} \sinh^2({g
\Delta \tau \over 2} - i g \epsilon)}\right\rbrace ^{-1}\eqn\qgfnaa.$$
Using the expansion

$$\csc^2 {\pi x}\; =\; {\pi}^{-2} \sum_{n=-\infty}^{\infty} (x\; -\; n)^{-2}
\eqn\qq$$
we can express \qgfnaa\ as

$$G_{acc}^{+}(\Delta \tau) = \left({-1 \over 4 \pi^2}\right) \sum_{n=-\infty}
^\infty {(\Delta \tau - 2 i \epsilon + 2 \pi i g^{-1} n)}^{-2}\eqn\qgfnab.$$
Substituting \qgfnab\ into \qprobrate\ and performing the contour integral
in the lower-half of the complex plane we obtain the transition probability
rate of the detector to be

$${\cal R}_{acc}(\Omega)\; =\; \left({1 \over 2 \pi}\right) \sum_{\Omega}
{{\vert {\cal M} \vert}^2\; \Omega \over {(e^{2 \pi g^{-1} \Omega}-1)}}
\eqn\qq$$
which is the well known thermal spectrum.
\smallskip
Having reviewed the Unruh-DeWitt detector theory and calculated the
transition probability rate of the detectors for the whole history of the
detector trajectory, {\it i.e} for infinite time intervals, let us now get
on with the crux of this paper, {\it viz} finite time detectors.
\bigskip
\beginsection{\bf 3. Detector response with window functions}

To understand some of the subtlities mentioned above, we shall begin with a
simple example.

Consider an Unruh-DeWitt detector which was moving in a trajectory $x(\tau)$
and was switched on during the interval $\tau = -T$ to $\tau = T$. The
response of such a detector is governed by the integral

$${\cal F}(\Omega, T)\; =\; \int_{-T}^T d\tau\; \int_{-T}^T d\tau'\; e^{-i
\Omega (\tau-\tau')}\; G^{+}(\tau), \tau')\eqn\aa.$$
We shall further assume that the trajectory is along the integral curve to
a timelike Killing vector field so that $G^{+}(\tau, \tau')\; =\; G^{+}
(\tau-\tau')$. {\it It is clear from this expression that ${\cal F} \rightarrow
0$ as $T \rightarrow 0$ irrespective of any other details}. Similarly, we
should recover the standard results when $T \rightarrow \infty$.
\smallskip
We shall now rewrite this expression differently and take the limits $T
\rightarrow 0$ and $T \rightarrow \infty$. Changing the variables to

$$x\; =\; \tau-\tau';\qquad y\; =\; \tau+\tau'\eqn\qtransfo$$
so that

$$\int_{-T}^T d\tau\; \int_{-T}^T d\tau'\; e^{-i \Omega (\tau -\tau')}\;
G^{+}(\tau- \tau')\; =\; \left({1 \over 2} \right)\; \int_{-2T}^{2T}dx\;
\int_{-2T+{\vert x \vert}}^{2T-{\vert v \vert}}dy\; e^{-i \Omega x}\;
G^{+}(x)\eqn\qq.$$
The factor $1/2$ is the Jacobian of the transformation from the $(\tau,
\tau')$ coordinates to the $(x, y)$ coordinates. Using this, we get

$${\cal F}(\Omega, T)\; =\;\int_{-2T}^{2T}dx\; e^{-i \Omega x}\;G^{+}(x)\;
(2T\;  -\; \vert x \vert)\eqn\cc.$$
\smallskip
Let us now consider the limits $T \rightarrow \infty$ and $T \rightarrow 0$
of this integral. When $T \rightarrow \infty$, we get

$${\cal F}(\Omega, T)\;= \;\lim_{T \rightarrow \infty}\left\lbrace(2T)\;
{\tilde G}^{+}(\Omega)\;-\;\int_{-2T}^{2T} dx\; e^{-i \Omega x}\;G^{+}(x)
\;\vert x \vert \right\rbrace\eqn\qq$$
where ${\tilde G}^{+}(\Omega)$ is the Fourier transform of $G^{+}(x)$.
Clearly

$$\eqalignno{{\cal R}(\Omega)\; =\;\lim_{T \rightarrow \infty}\left\lbrace{
{\cal F}(\Omega, T) \over {2T}}\right\rbrace\;& = \; \lim_{T \rightarrow
\infty} \left\lbrace {\tilde G}^{+}(\Omega)\; -\;\left({1 \over 2T}\right)
\int_{-\infty}^{\infty}dx\;  e^{-i \Omega x}\;G^{+}(x)\;\vert x \vert
\right\rbrace\cr
&=\; {\tilde G}^{+}(\Omega)&\eqnx\qq\cr}$$
provided the second integral is well defined. This expression is finite and
represents a constant rate of transition; we have thus recovered the
standard result in the necessary limit.
\smallskip
Let us next consider the $T \rightarrow 0$ limit which is somewhat tricky.
We need to evaluate

$${\cal F}(\Omega, 0)\; =\; \lim_{T \rightarrow 0}\; \int_{-2T}^{2T}
dx\;e^{-i \Omega x}\; G^{+}(x)\; (2T - \vert x \vert)\eqn\xx.$$
The integral over $x$ is confined to a small range $(-2T, 2T)$ around the
origin. This implies that we can expand the integrand in a Taylor series
around the origin to obtain the required limit. We write

$$e^{-i \Omega x}\; G^{+}(x)\; \simeq\; (1-i\Omega x-{1 \over 2}{\Omega}^2
x^2+ ....)\; (G^{+}(0) +G'^{+}(0)\; x+{1 \over 2}\; G''^{+}(0)\; x^2 +
....)\eqn\qq.$$
Substituting the above expression into \xx\ and integrating we obtain

$$\eqalignno{{\cal F}(\Omega, T)\; &\simeq\; 4T^2 G^{+}(0)\; -\; {4 \over
3}T^4\; (G''^{+}(0)-{\Omega}^2 G^{+}(0)-2i \Omega G'^{+}(0))\; +\; O(T^4{
\omega}^4)\cr
&\simeq \; 4 T^2\; G^{+}(0)&\eqnx\qq\cr}$$
to the lowest order. All derivatives of $G^{+}(x)$ in (3+1) dimensions behave
as ${\epsilon}^{-n}$ at origin and in particular,

$$G^{+}(0)\; =\; {1 \over {4{\pi}^2 {\epsilon}^2}}\eqn\qq$$
giving

$${\cal F}(\Omega, T)\; \simeq\; \left({T^2 \over {{\pi}^2{\epsilon}^2}}
\right)\eqn\bb.$$
\smallskip
The above expression shows that care should be excercised when the limits
$T \rightarrow 0$, $\epsilon \rightarrow 0$ are incorporated. It is clear
from the fundamental definition of the integral in \aa\ that we must have
${\cal F}(\Omega, 0)\; =\; 0$ for all regular integrands. If the integrand
has a pole in the real axis (requiring an $i\epsilon$ prescription to give
meaning to the integral) then we should {\it arrange} the limiting procedure
in such a way that ${\cal F}(\Omega, 0)\; =\; 0$. This can be achieved by
using the rule that $\epsilon \rightarrow 0$ limit should be taken right at
the end, after the limit $T \rightarrow 0$ has been taken. Since

$$\lim_{\epsilon \rightarrow 0} \biggl\lbrace \lim_{T \rightarrow 0} \;
\left({T^2 \over {{\epsilon}^2}}\right)\biggl\rbrace\;=\; 0;\qquad \lim_{T
\rightarrow 0}\biggl\lbrace\lim_{\epsilon \rightarrow 0}\; \left({T^2 \over
{{\epsilon}^2}}\right) \biggl\rbrace\; =\; \infty\eqn\qq$$
only the former ordering will provide physically reasonable results. This
prescription is also necessary to ensure that $G^{+}(0), G'^{+}(0)...$etc
exist in the Taylor expansion for $G^{+}(x)$. For $\epsilon = 0$, this
expansion ceases to exist.
\smallskip
In (1+1) dimensions $G^{+}(x)$ has a logarithmic dependence in $x$; hence
the limit will be modified to the form

$${\cal F}(\Omega, T)\; \propto\; T^2\; \ln({\epsilon}^2)\eqn\qq.$$
taking $T \rightarrow 0$ limit first will give the sensible result ${\cal F}
(\Omega, 0)\; =\; 0$ while if $\epsilon \rightarrow 0$ limit is taken first
we will obtain a logarithmic divergence. We shall see explicit examples of
such ambiguities (and their resolution) in what follows.
\smallskip
We shall now calculate the response of three different kinds of detectors
which have been switched on for a finite time. In selecting these examples,
we are motivated by the fact that no realistic detector can be switched on
abruptly. Hence, instead of working with \aa\ , we will consider the
integral of the form

$${\cal F}(\Omega, T)\; =\; \int_{-\infty}^\infty d\tau\; \int_{-\infty}^
\infty d\tau'\; e^{-i \Omega (\tau-\tau')}\; W(\tau, T)\; W(\tau', T)\;
G^{+}(x(\tau), x(\tau'))$$
where $W(\tau, T)$ is a ``window function" with the properties

$$W(\tau, T)\; =\; \cases{1 & (for $-T \ll \tau \ll T)$ \cr 0 & (for $\vert
\tau \vert\gg T$) \cr}\eqn\qq.$$
The abrupt switching corrseponds to $W(\tau, T)\; =\; \Theta(T-\tau)\; +\;
\Theta(T+\tau)$. More gradual switching on and off can be mimicked, for
{\it e. g}, by the functions

$$W_{1}(\tau, T)\; =\; \exp(-{\tau^2 \over {2T^2}})\; ;\qquad\; W_{2}(\tau,
T)\; =\; \exp(- {\vert \tau \vert \over T})\eqn\qq$$
etc. In order to see the effects of smoothness of the window functions on
the detector response, we shall discuss the results for all the three cases:
$W_1, W_2$ and $W$, in that order.
\smallskip
The motivation to study the detector response with smooth window functions
$W_1$ and $W_2$ is to carefully identify any possible divergence that may
arise when a finite time detection is performed. We study the detector
response with a gaussian window function ($W_1$) in {\bf 3(a)}, in the
presence of a window function with an exponential cut-off ($W_2$) in {\bf
3(b)} and show that no divergences arise in these cases. In the third sub-
section {\bf 3(c)} we calculate the response for the window function $W$.
All these results remain finite if the limits are handled carefully.
\medskip
\centerline{\bf 3(a). Gaussian window function:}
\noindent
The detector response integral with the window function $W_1$ is

$${\cal F}(\Omega, T)\; =\; \int_{-\infty}^\infty d\tau\; \int_{-\infty}^
\infty d\tau'\; e^{-i \Omega (\tau-\tau')}\; exp(-{\tau \over T}^2)\;
exp(-{\tau' \over T}^2)\; G^{+}(\tau, \tau')\eqn\qq$$
which can be rewritten as

$${\cal F}(\Omega, T)\; =\; \int_{-\infty}^\infty d\tau \int_{-\infty}^\infty
d\tau' \; e^{-i \Omega (\tau-\tau')}\; G^{+}(x(\tau), x(\tau'))\; \exp-{1
\over {2 T^2}}\left((\tau + \tau')^2 + (\tau - \tau')^2\right)\eqn
\qdetresggfnt.$$
Substituting the Wightman function \qgfni\ for the inertial trajectory
in the above integral and performing the transformations \qtransfo\ the
integral simplifies to

$${\cal F}_{ine}(\Omega, T)\; =\; \left({1 \over 2}\right) \int_{-\infty}
^{\infty}dy\; \exp(-{y^2 \over {2T^2}}) \int_{-\infty}^{\infty}dx\; \left
\lbrace{-1 \over 4 \pi^2 (x -i\epsilon)^2} \right\rbrace\; e^{-i \Omega x}
\; \exp(-{x^2 \over {2 T^2}})$$
$$=\; -{T \over {8\pi^2}} {\sqrt {2 \pi}}\; I \eqn\qq$$
where

$$I\; =\; \int_{-\infty}^{\infty}{dx \over {(x -i\epsilon)}^2}\; e^{-i
\Omega x}  \;exp(-{x^2 \over {2 T^2}})\eqn\qint.$$
Writing the gaussian function in $x$ in the above integral as a Fourier
transform using

$$\exp(-{x^2 \over {2 T^2}})\; =\; {T \over {\sqrt{2 \pi}}}\; \int_{-
\infty}^{\infty} dk\; \exp(-{k^2 T^2 \over 2})\; e^{ikx}\eqn\qgft$$
and interchanging the order of integration we obtain

$$I\; =\; {T \over {\sqrt{2 \pi}}} \int_{-\infty}^\infty dk \; \exp(-{k^2
T^2\over 2}) \int_{-\infty}^{\infty} dx \; {e^{i(k-\Omega)x} \over (x -i
\epsilon)^2} \eqn\qq.$$
When $k>\Omega$, the $x$ integral can be performed as a contour integral
by closing the contour in the upper half of the complex $x$-plane and the
second order pole at $x=i\epsilon$ gives the non-trivial contribution to
the integral. But, when $k<\Omega$ the contour has to be closed in the
lower-half and  since the function is analytic in this half the integral
vanishes. Hence the limits of the $k$-integral may be set to $\Omega$ and
$\infty$. After some manipulations and substituting this result in \qint\ ,
we get

$${\cal F}_{ine}(\Omega, T)\; =\; \left\lbrace{\exp({\epsilon}^2/ {2 T^2})\;
\exp(\Omega \epsilon) \over {2\; \pi}}\right\rbrace \int_r^\infty dp \;
e^{-p^2} (p-r)\eqn\qdetresit$$
where

$$p\; =\; {k T \over {\sqrt 2}} + {\epsilon \over {{\sqrt 2} T}}; \qquad
r\; =\; {\Omega T \over {\sqrt 2}} + {\epsilon \over {{\sqrt 2}T}}\eqn
\qq.$$
\smallskip
Before proceeding further let us check that the expression \qdetresit\
gives sensible results for the limits $T \rightarrow 0$ and $T \rightarrow
\infty$. Since this is an inertial detector we must have ${\cal F}(\Omega,
\infty)\; =\; 0$ and for a detector on any trajectory we should have ${\cal
F}(\Omega, 0)\; =\; 0$. These two limits can be obtained from the above
result. In the $T \rightarrow \infty$ the lower and the upper limits of
the above integral coincide the integral vanishes identically, thus
reproducing the result appropriate for the inertial detector. (Note that
for large $r$,

$$r \int_r^{\infty}dp\; e^{-p^2} \; \simeq \; {1 \over 2} e^{-{r^2}}
\left\lbrace 1 + O({1 \over {r^2}})\right\rbrace\eqn\qq$$
vanishes exponentially). Hence, there is no ambiguity in this result.
\smallskip
Studying the limit $T \rightarrow 0$ of \qdetresit\ , when the window
function is sharply peaked at the origin, has to be done more carefully
as follows. In this case, because it matters whether the limit $T \rightarrow
0$ is taken first and the condition $\epsilon \rightarrow 0$ is incorporated
later or vice-versa. The earlier alternative is to be adopted (as has been
mentioned earlier) for the reason that the $\epsilon$ term helps us to
identify the poles in the contour integrals; hence unless and until all
the other limits in the problem have already been taken care of, the limit
on $\epsilon$ should not be incorporated. Keeping this point in mind, we
consider the limit  $T \rightarrow 0$, $r\rightarrow (\epsilon/ {{\sqrt 2}T})
$ and rewrite ${\cal F'}_{ine}(\Omega, T))$ as

$${\cal F}_{ine}(\Omega, T)\; =\; \left\lbrace{\exp({\epsilon}^2/ {2 T^2})
\; \exp(\Omega \epsilon) \over {2\; \pi}}\right\rbrace\; B$$
where

$$B\; =\; \left\lbrace \int_{\epsilon \over {{\sqrt 2}T}}^{\infty} dp \;
e^{-p^2} p -{\epsilon \over {{\sqrt 2}T}}\left(\int_0^\infty dp \; e^{-p^2}
- \int_0^{\epsilon \over {{\sqrt 2}T}} dp \; e^{-p^2}\right)\right\rbrace
\eqn\qdetresiti.$$
The last term in the above expression is the error function and its
asymptotic form for large arguments is given to be

$${2 \over {\sqrt {\pi}}} \int_0^x dv \; e^{-v^2}\; =\; 1 - \; {e^{-x^2}
\over {\sqrt \pi}}\; \left\lbrace {1\over x}-{1 \over 2 x^3}+{3 \over
4x^5}.......\right\rbrace\eqn\qq.$$
Substituting the above expression in \qdetresiti\ , we obtain the detector
response when $T \rightarrow 0$ to be

$${\cal F}_{ine}(\Omega, 0)\; =\; \left({e^{\Omega \epsilon} \over {4 \pi}}
\; {T^2 \over {\epsilon^2}}\right) \rightarrow 0 \eqn\qq$$
for finite $\epsilon$. This expression has the same form as \bb\ and
clearly illustrates the need to keep $\epsilon \neq 0$ till the end. Note
that the detector response function as well the rate of transition ${\cal
R}_{ine}(\Omega, T) \propto {\cal F}_{ine}(\Omega, T)/ T$ vanish when $T
\rightarrow 0$. The non-commutativity of the limiting procedure as regards
$T \rightarrow 0$, $\epsilon \rightarrow 0$ in the detector response
functions is evident due to the presence of factors like ${\epsilon/ T}$.
If the condition $\epsilon \rightarrow 0$ is incorporated first in
\qdetresit\ it factorises to

$${\cal F'}_{ine}(\Omega, 0)\; =\; \left({1 \over \pi}\right)\; \int_{
\Omega T \over {\sqrt 2}}^{\infty}dp\; e^{-p^2}\; \left(p - {\Omega T
\over {\sqrt 2}}\right)\eqn\qq.$$
If we now take the limit $T \rightarrow 0$ we obtain

$${\cal F'}_{ine}(\Omega, 0)\; =\; ({1 \over \pi})\; \int_0^{\infty}dp\;
e^{-p^2}\; p\; =\; {1 \over {2 \pi}}\eqn\qq.$$
On the other hand, ${\cal F}_{ine}(\Omega, T)$ vanishes if we take the
limit $T \rightarrow 0$ before we set $\epsilon\; =\; 0$. We stress again
that the procedure of letting $\epsilon$ to zero only after the $T
\rightarrow 0$ limit is taken is the proper one.
\smallskip
If we are only interested in finite, non-zero values of $T$, then we can
set $\epsilon \; =\; 0$ in the integral \qdetresit\ can be written in a
closed form as

$${\cal F}_{ine}(\Omega, T)\; =\; {1 \over {4 \pi}}\;\left\lbrace \exp-
({{\Omega}^2 T^2 \over 2})\;-\;\left({\Omega T \over {\sqrt 2}}\right)\;
\Gamma\left({1 \over 2}, {{\Omega}^2 T^2 \over 2}\right)\right\rbrace\eqn\qq.$$
                    For $\Omega T \gg 1$, this expression has the asymptotic
form

$${\cal F}_{ine}(\Omega, T)\; \simeq\; {1 \over {4 \pi}}\; {\exp(-{\Omega}^2
T^2/ 2)\over {{\Omega}^2 T^2}}\eqn\qq.$$
This shows that an inertial detector, switched on for a finite period of time,
does give a non-zero response which goes to zero exponentially as $T
\rightarrow
\infty$.
\smallskip
Let us now carry out the same analysis for the case of an accelerated detector.
For this case, the Wightman function given by \qgfnab\ , when substituted into
\qdetresggfnt\ and the transfomations \qtransfo\ when incorporated, the result
is

$${\cal F}_{acc}(\Omega, T)\; =\; -\left({1 \over {8 \pi^2}}\right) \int_
{-\infty}^{\infty}dy\; \exp(-{y^2 \over {2 T^2}}) \int_{-\infty}^{\infty}dx\;
\left \lbrace \sum_{n=-\infty}^\infty {e^{-i \Omega x} \; \exp(-{x^2/{2 T^2}})
\over (x -i b_n)^2} \right\rbrace\eqn\qq$$
where $b_n\; =\; \epsilon- 2 \pi g^{-1}n$. With the aid of \qgft\ , the above
integral can be simplified to the form

$${\cal F}_{acc}(\Omega, T)\; =\; -{T \over {8 \pi^2}} \; {\sqrt {2 \pi}}
\sum_{n={-\infty}}^\infty I_n\eqn\qq$$
where

$$I_n\; =\; {T \over {\sqrt{2 \pi}}} \int_{-\infty}^{\infty} dk \; \exp({-k^2
T^2 \over2})\; \int_{-\infty}^{\infty}dx\; {e^{i(k-\Omega) x} \over {(x-ib_n)
^2}}\eqn\qq.$$
When $k>\Omega$ the $x$ integration has to be performed by closing the contour
in the upper-half of the complex $x$-plane and the poles corresponding to the
values of $n$ between $-\infty$ and zero contribute nontrivially to ${\cal
F}_{acc}(\Omega, T)$ giving,

$${\cal F}_{acc1}(\Omega, T)\; =\; \sum_{n=-\infty}^0 \left\lbrace{\exp({b_n}
^2/ {2 T^2}) \; \exp(\Omega b_n) \over {2\; \pi}}\right\rbrace \int_{r'}^
\infty dp'\; e^{-p'^2} \; (p'- r')\eqn\qq$$
where

$$p'\; =\; {k T \over {\sqrt 2}} +{b_n \over {{\sqrt 2}T}};\qquad r'\; =\;
{\Omega T \over {\sqrt 2}}+ {b_n \over {\sqrt 2}T}\eqn\qq.$$
When $k<\Omega$ the contour has to be closed in the lower half plane with the
contributions arising from the poles corresponding to the values of $n=1, 2,
3, ....$:

$${\cal F}_{acc2}(\Omega, T)\; =\; \sum _{n=1}^{\infty} \left({\exp({b_n^2/ 2
T^2}) \; \exp({\Omega b_n}) \over {2\; \pi}}\right) \int_{-r'}^\infty dp'\;
e^{-p'^2} \; (p'+r')\eqn\qq.$$
The complete result is ${\cal F}_{acc}(\Omega, T)\; =\; {\cal F}_{acc1}(\Omega,
T)\; +\; {\cal F}_{acc2}(\Omega, T)$, {\it i.e}

$${\cal F}_{acc}(\Omega, T)\; = \; \sum_{n=-\infty}^0 \left\lbrace{{\exp({b_n}
^2/ {2 T^2})\; \exp(\Omega b_n)} \over {2\; \pi}}\right\rbrace \int_{r'}^{
\infty} dp' \; e^{-p'^2} (p'- r')$$
$$+ \; \sum_{n=1}^{\infty} \left({{e^{{b_n}^2 \over {2 T^2}} \; e^{\Omega
b_n}} \over {2\; \pi}}\right\rbrace \int_{-r'}^{\infty} dp'\; e^{-p'^2}\;
(p'+ r')\eqn\qdetresgat.$$
\smallskip
Let us again check the two relevant limits. In the limit $T \rightarrow
\infty$ the lower limit of the above integrals reduce to $\infty$ and
$-\infty$ respectively, so that only ${\cal F'}_{acc2}(\Omega)$ contributes
to the detector response. Evaluating this and imposing the condition $\epsilon
\rightarrow 0$, we get the standard result:

$${\cal F}_{acc}(\Omega)\; =\; {T \over {2\; \sqrt{2 \pi}}}\; {\Omega \over
{(e^{2 \pi g^{-1} \Omega} - 1})}\eqn\qq.$$
In this case the ratio ${\cal R}_(acc)\; =\; {{\cal F}_{acc}(\Omega)/ T}$
should be intrepreted as the transition probability rate.
\smallskip
When $T\rightarrow 0$, we can perform the analysis as in the case of inertial
detector, since only the $n=0$ term in the series \qdetresgat\ contributes
nontrivially; we obtain the result to be

$${\cal F}_{acc}(\Omega, 0)\; =\; {e^{\Omega \epsilon} \over {2 \pi}}\;
{T^2 \over {\epsilon}^2}\eqn\qq.$$
This is identical to the inertial detector result and shows that the
transition probablity (as well the rate) will go to zero as $T \rightarrow
0$.
\smallskip
The fact that both accelerated and inertial detectors give identical results
for the $T \rightarrow 0$ limit is to be expected on physical grounds. The
curvature of the trajectory cannot make its presence felt for infinitesimal
intervals and the response of the detector cannot depend on parameters like
$g$ which charecterise the detector trajectory.
\smallskip
(Note that, for any T, the detection is now due to two effects: (i) The
trajectory being non-inertial and (ii)the detector being kept switched on
only for a finite time. The second effect is present even for inertial
trajectory. It may be physically more useful to subtract the inertial
response from the accelerated detector response to obtain the effects that
are uniquely due to (i). In this case ${\cal F}_{net}\; =\; {\cal F}_{acc}\;
-\; {\cal F}_{ine}$ vanishes trivially for $T \rightarrow 0$. This is, of
course, not mandatory to obtain sensible results.)
\smallskip
It is possible to state some of these results in greater generality for this
window function. Note that for a detector moving along any trajectory for
which $G^{+}(x, x')\; =\; G^{+}(\tau-\tau')$ the response function is

$$\eqalignno{{\cal F}(\Omega,T)\; &=\; \int_{-\infty}^{\infty} d\tau \int_
{-\infty}^{\infty} d\tau' \exp-{1\over {2T^2}}({\tau}^2 + {\tau'}^2)\;
e^{-i \Omega (\tau-\tau')} G^{+}(\tau-\tau')\cr
&=\; \left({1 \over 2}\right)\; \int_{-\infty}^{\infty}dy\; \exp(-{y^2
\over {2T^2}})\; \int_{-\infty}^{\infty} dx\; \exp(-{x^2 \over {2T^2}})
\; e^{-i\Omega x}\; G^{+}(x)\cr
&=\; T\; \sqrt{\pi \over 2}\; \int_{-\infty}^{\infty}dx\; \exp(-{x^2
\over {2T^2}})\; e^{-i\Omega x}\; G^{+}(x)&\eqnx\qone.\cr}$$
We can write

$$f(x)\; \left( e^{-i\Omega x}\; G^{+}(x) \right)\; =\; f\left( i{\partial
\over {\partial \Omega}} \right) \left[ e^{-i\Omega x}\; G^{+}(x) \right]
\eqn\qq$$
for any function $f(x)$ which has a power series expansion around $x\;=\;
0$. Hence we can write

$$\eqalignno{{\cal F}(\Omega, T)\;&=\; T\; \sqrt{\pi \over 2}\; \int_
{-\infty}^{\infty}dx\; \left\lbrace \exp\left({1 \over T}\; {\partial^2
\over {\partial {\Omega}^2}}\right) \right\rbrace \left[  e^{-i\Omega x}\;
G^{+}(x) \right]\cr
&=\; \exp\left({1 \over {2T^2}}\; {\partial^2 \over {\partial {\Omega}^2}}
\right)\; \left[{\cal F}(\Omega, \infty)\right]& \eqnx\qq.\cr}$$
The expression in the square brackets is the result for the infinite time
detector. The corresponding rates are

$${\cal R}(\Omega, T)\; =\; \exp\left({1 \over {2T^2}}\; {\partial^2 \over
{\partial {\Omega}^2}}\right)\; \left[{\cal R}(\Omega, \infty)\right]\eqn
\qq.$$
This formula allows us to systematically calculate finite time corrections
as a series in $T^{-1}$. To the lowest order, the correction is

$${\cal R}(\Omega, T)\; =\; {\cal R}(\Omega, \infty)\; +\; ({1 \over {2T^2}}
)\; {{\partial}^2 \over {\partial {\Omega}^2}}\; {\cal R}(\Omega, \infty)\;
+\; O(T^{-4})\eqn\qtwo .$$
In the case of uniformly accelerated detector, we get

$${\cal R}(\Omega, T)\; \simeq\; {\cal R}(\Omega, \infty)\; \left\lbrace{1
- {1 \over T^2} {2\pi \over {g\Omega}} {e^{2\pi\Omega g^{-1}} \over {(e^{2
\pi\Omega g^{-1}} -1)}^2} \left(e^{2\pi\Omega g^{-1}} -1 -\pi\Omega g^{-1}
(e^{2\pi\Omega g^{-1}}+1)\right)}\right\rbrace\eqn\qq.$$
\medskip
\centerline{\bf 3(b). Window function with an exponential cut-off}
\noindent
Having studied the detector response with a gaussian window function, we now
study the same with the window function of the type $W_2$. In this case the
response function turns out to be

$${\cal F}(\Omega, T)\; =\; \int_{-\infty}^\infty d\tau \int_{-\infty}^\infty
d\tau' \; e^{-i \Omega (\tau-\tau')} \; \exp-{1 \over T}(\vert \tau \vert
\; +\; \vert \tau'\vert)\; G^{+}{(x(\tau), x(\tau'))}\eqn\qq.$$
Introducing the window functions as Fourier transforms

$$\exp-({\vert \tau \vert \over T})\; =\; \int_{-\infty}^{\infty} dk\; f(k)
\; e^{ik \tau};\qquad f(k)\; =\; {T \over {\pi}}\; {1 \over (1+k^2T^2)}\eqn
\qeft$$
and performing the tranformations \qtransfo\ we obtain the detector response
for the case of an inertial detector to be

$${\cal F}_{ine}(\Omega, T)\; =\; (-{1 \over {8 {\pi}^2}})\; \int_{-\infty}
^\infty dk\; f(k) \int_{-\infty}^\infty dq\; f(q) \int_{-\infty}^\infty dy
\; e^{i{y\over 2}(k+q)} \int_{-\infty}^\infty dx \; {e^{ix({k-q\over 2} -
\Omega )} \over {(x-i \epsilon)}^2}\eqn\qq.$$
When the $y$ and the $q$ integrals in the above expression are performed
in that order, the result is

$${\cal F}_{ine}(\Omega, T)\; =\; (-{1\over {2\pi}}) \int_{-\infty}^\infty
dk\; f(k)\; f(-k) \int_{-\infty}^{\infty}dx\; {e^{i(k-\Omega)x} \over
{(x-i\epsilon)}^2}\eqn\qq.$$
Performing the contour integral after substituting for $f(k)$, the detector
response function reduces to

$${\cal F}_{ine}(\Omega, T)\; =\; ({1 \over {\pi}^2})\; e^{\Omega \epsilon}
\int_{\Omega T}^\infty dp\; {\exp(-{p \epsilon/ T}) \over {(1+p^2)}^2}\; (p
-\Omega T)\eqn\qdetreseit$$
where $p=kT$. When $T\rightarrow \infty$ the limits of the above integral
coincide giving a null result as expected. When $T\rightarrow 0$ the lower
limit of the above integral goes to zero and so does the second term in the
integrand with the result

$${\cal F}_{ine}(\Omega, 0)\; =\; ({1 \over {\pi}^2})\; e^{\Omega \epsilon}
\int_0^\infty dp \; {p \; \exp(-{p\epsilon / T}) \over {(1+p^2)}^2}\eqn\qq$$
which reduces to zero in the limit $T \rightarrow 0$ being exponentially
damped out by the $\exp(-{p \epsilon/ T})$ factor. We again note the crucial
role played by the $\epsilon$ factor. The limits $\epsilon \rightarrow 0$,
$T \rightarrow 0$ do not(again!) commute in the function $\exp(-{p \epsilon/
T})$:

$$\lim_{T \rightarrow 0}\left\lbrace\lim_{\epsilon \rightarrow 0} \exp(-p\;
\epsilon/ T)\right\rbrace\;=\;1;\qquad \lim_{\epsilon \rightarrow 0}\left
\lbrace\lim_{T \rightarrow 0}\; \exp(-p\; \epsilon/ T) \right\rbrace\;=\;0
\eqn\qq.$$
Sensible result for the inertial detector is obtained with the latter
sequence, as we have emphasised several times by now.
\smallskip
If we are interested only in the $T \neq 0$ case, then we can set $\epsilon
= 0$ in \qdetreseit\ ; this integral with $\epsilon = 0$ can be expressed in
closed form:

$${\cal F}_{ine}(\Omega, T)\; =\;\left({1 \over {2{\pi}^2}}\right)\; \left
\lbrace{1 \over {(1+{\Omega}^2T^2)}}\; -\; {\Omega T \over 2}\left\lbrace\pi
\;-\;2 \tan^{-1}(\Omega T)\;-\;\sin2(\tan^{-1}(\Omega T))\right\rbrace\right
\rbrace\eqn\qq.$$
For $\Omega T \gg 1$, this function behaves as

$${\cal F}_{ine}(\Omega, T)\; \simeq\; {1 \over {6 {\pi}^2}}\; {1 \over
{\Omega^2 T^2}}\eqn\qq.$$
We once again see that the inertial detector will respond in the Minkoski
vacuum if it is kept switched on only for a finite $T$. As $T \rightarrow
\infty$, this response dies as $T^{-2}$. For the case of an accelerated
detector the response function, the corresponding  integral factorises to
be

$${\cal F}_{acc}(\Omega, T)=(-{1 \over 8{\pi^2}}) \sum_{n=-\infty}^\infty
\int_{-\infty}^\infty dk\; f(k) \int_{-\infty}^\infty dq\; f(q) \int_
{-\infty}^\infty dy\; e^{i{y\over 2}(k+q)} \int_{-\infty}^\infty dx\;
{{e^{i({k-q \over 2}-\Omega)x}
\over {(x-ib_n)}^2}}\eqn\qq$$
where $b_n = \epsilon - 2\pi g^{-1} n.$ Performing the $y$ and the $q$
integrals in that order the detector response function reduces to

$${\cal F}_{acc}(\Omega, T)\; =\; (-{1 \over {2\pi}}) \sum_{n=-\infty}^\infty
\int_{-\infty}^{\infty} dk\; f(k)\; f(-k) \int_{-\infty}^\infty dx\;
{e^{i(k-\Omega)} \over {(x-ib_n)}^2}\eqn\qq$$
The above contour integral can be performed as before to give the following
result:

$${\cal F}_{acc}(\Omega, T)\; =\; ({1 \over {\pi}^2}) \; \biggl\lbrace
\sum_{n=-\infty}^0 e^{\Omega b_n} \int_{\Omega T}^\infty dp\; {e^{-p b_n
\over T} \over {(1+p^2)}^2} \; (p -\Omega T)$$
$$+\; \sum_{n=1}^{\infty}\; e^{\Omega b_n} \int_{-\Omega T}^\infty dp\;
{e^{p b_n \over T} \over {(1+p^2)}^2}\; (p +\Omega T)\biggl\rbrace\eqn\qq$$
where $p=kT$. When $T \rightarrow \infty$, the $\exp(-p\; b_n/ T)$ factors
in the integrand reduce to unity and the lower limit of the integrals are
$\infty$ and $-\infty$ respectively. As the limits coincide the first
integral vanishes. In the second integral only the second term contributes,
the first term being an odd function reducing to zero in the symmetric
limits. Thus, in the $T \rightarrow \infty$ limit we recover the Fulling-
Unruh-Davies thermal spectrum after $\epsilon \rightarrow 0$:

$${\cal F}_{acc}(\Omega)\; =\; {1 \over {\pi}^2}\; \Omega T \left\lbrace
\int_{-\infty}^{\infty} {dp \over {{(1+p^2)}^2}} \right\rbrace \; \left
\lbrace \sum_{n=1}^{\infty} e^{-2 \pi g^{-1} \Omega n} \right\rbrace = {T
\over 2 \pi} \; {\Omega \over {(e^{2 \pi \Omega g^{-1}} -1)}}\eqn\qq.$$
In this case, ${\cal F}_{acc}(\Omega)/ T$ is to be interpreted as the rate
of transition probablity of the detector.
\smallskip
When $T \rightarrow 0$ only the $n\;=\;0$ term contributes non-trivially
so that the response function factorises to

$${\cal F}_{acc}(\Omega, 0)\; =\; \left({1 \over \pi^2}\right)\; e^{\Omega
\epsilon}\int_0^\infty dp\; {p\; \exp(-p \epsilon/ T) \over {(1+p^2)}^2}
\eqn\qq$$
which vanishes when $T\rightarrow 0$ because of the exponential damping
factor in the integrand.
\medskip
\centerline{\bf 3(c). A rectangular window function (sum of two step-
functions)} \noindent
In this section we study the detector response for explicit finite time
limits without introducing smooth window functions. The detector response
integral for this case is given by \aa\ and when the transformations
\qtransfo\ are performed it reduces to \cc\ , that is

$${\cal F}(\Omega, T)\; =\; \int_{-2T}^{2T} dx \; e^{-i\Omega x} \; (2T-
\vert x \vert) \; G^+(x)\eqn\qq.$$
For the case of an inertial detector the integrals to be evaluated are

$${\cal F}_{ine1}(\Omega, T)\; =\; (-{2 T \over {4 {\pi}^2}}) \; \int_
{-2T}^{2T} dx \; {e^{-i \Omega x} \over {(x-i \epsilon)}^2}\eqn\qq$$
and

$${\cal F}_{ine2}(\Omega, T)\; =\; ({1 \over {4 \pi^2}}) \; \int_{-2T}^
{2T} dx \; {e^{-i \Omega x} \over {(x-i \epsilon)}^2} \; \vert x \vert
\eqn\qq$$
so that ${\cal F}_{ine}(\Omega, T)\; =\; {\cal F}_{ine1}(\Omega, T)\; +
\; {\cal F}_{ine2}(\Omega, T)$. This finite time response of the inertial
detector can be obtained by evaluating the above integrals. The detailed
calculations are given in Appendix A. The result is

$${\cal F}_{ine}(\Omega, T)\; =\; {1 \over {4 {\pi}^2}} \biggl\lbrace -
e^{2i \Omega T} \int_0^{\infty} dv\; {e^{-\Omega v} v \over {(v+ \epsilon
-2iT)}^2} - e^{-2i \Omega T} \int_0^{\infty}dv\; {e^{-\Omega v} v \over
{(v+ \epsilon +2iT)}^2}$$
$$+ 2 \int_0^{\infty} dv {e^{-\Omega v} v \over {(v+ \epsilon)}^2}\biggl
\rbrace\eqn\qdetressit.$$
\smallskip
The two limits again give sensible results. When $T \rightarrow 0$, the
first two integrals exactly cancel the third giving ${\cal F}_{ine}\; =\;
0$, provided we keep $\epsilon \neq 0$. If we set $\epsilon\; =\; 0$ before
we set $T\;=\;0$, then the limit $T \rightarrow 0$ will produce logarithmic
divergences at the lower limit of integration. For large $T$, the rate
${\cal R}_{ine}\; =\;{{\cal F}_{ine}/ T}$ vanishes because $ {\cal F}_
{ine}$ is bounded and well defined in this limit while $T \rightarrow
\infty$.
\smallskip
For the accelerated detector case, the evaluation of the response integrals
is similar but a bit more involved. The response function is

$${\cal F}_{acc}(\Omega, T)\; =\; {-1 \over {4 {\pi}^2}} \sum_{n=-\infty}
^{\infty} \int_{-T}^T d\tau' \int_{-T}^T d\tau\; {e^{-i\Omega (\tau-\tau')
} \over {(\tau -\tau' -ib_n)}^2}\eqn\qq$$
where $b_n\; =\; \epsilon- 2 \pi g^{-1}n$. Performing the transformations
\qtransfo\ we obtain the response function to be

$${\cal F}_{acc}(\Omega, T)\; =\; \sum_{n=-\infty}^{\infty} \left\lbrace
{\cal F}_{a1n}(\Omega, T) + {\cal F}_{a2n}(\Omega, T)\right\rbrace\eqn\qq$$
where

$${\cal F}_{acc1n}(\Omega, T)\; =\; {-2T \over {4 \pi^2}} \int_{-2T}^{2T}
dx \; {e^{-i \Omega x} \over {(x-ib_n)}^2}\eqn\qq$$
and

$${\cal F}_{acc2n}(\Omega, T)\; =\;{1 \over {4 \pi^2}} \int_{-2T}^{2T} dx
\; {e^{-i \Omega x} \; \vert x \vert \over {(x-ib_n)}^2}\eqn\qq.$$
The finite time response of an accelerated detector can be obtained by
evaluating the above integrals. The calculation is given in Appendix B;
the result is

$${\cal F}_{acc}(\Omega, T)\; =\; {1 \over 4\pi^2} \sum_{n=-\infty}^{\infty}
\biggl\lbrace 2\pi\; 2T\; \Omega\; \Theta(n)\; e^{\Omega b_n} -  e^{2i
\Omega T} \int_0^{\infty} dv\; {e^{-\Omega v} v \over {(v+b_n -2iT)}^2}$$
$$ - e^{-2i \Omega T} \int_0^{\infty} dv\; {e^{-\Omega v} v \over {(v+b_n
+2iT)}^2} + 2 \int_0^{\infty} dv\; {e^{-\Omega v} v \over {(v+b_n)}^2}
\biggl\rbrace\eqn\qdetressat.$$
\smallskip
In the limit $T \rightarrow 0$ the above quantity reduces to zero, the first
term identically zero being proportional to T; the second and the third
terms being cancelled by the fourth one. Whereas in the infinite time limit,
concentrating on the transition probablity rate we obtain

$${\cal R}_{acc}(\Omega\; =\; {{\cal F}_{acc}(\Omega) \over (2T)}\; =\; {1
\over 4\pi^2}\;  \sum_{n=-\infty}^{\infty} 2\pi \; \Omega \; \Theta(n)\;
e^{\Omega b_n}\; =\; {1 \over 2 \pi}\; {\Omega \over {(e^{2 \pi \Omega
g^{-1}} - 1)}}\eqn\qq$$
a thermal spectrum, the other terms in \qdetressat\ vanishing when divided
by the infinite time interval.
\smallskip
Finally we shall provide an asymptotic formula for the detector response
with an arbitrary window function of the form $W(\tau/T)$. This is a
direct generalisation of the results in \qone\ to \qtwo\ . For a general
window function we can write

$$\eqalignno{{\cal F}(\Omega, T)\; & =\; \int_{-\infty}^{\infty}d\tau\;
\int_{-\infty}^{\infty}d\tau' \; W(\tau, T)\; W(\tau', T)\; e^{-i\Omega
(\tau-\tau')}\; G^{+}(\tau-\tau') \cr
& =\;W\left(i{\partial \over {\partial \Omega}}, T\right)\; W\left(-i{
\partial \over {\partial \Omega}}, T\right)\;{\cal F}(\Omega, T)&\eqnx
\qq.\cr}$$
Assuming that $W({\tau, T})\;=\;W({\tau \over T})$ has the Taylor
expansion

$$\eqalignno{W({\tau \over T})\;& \simeq\;W(0)\;+\;W'(0)\;({\tau \over
T})\;+\;{1 \over 2}\;W''(0)\;{({\tau \over T})}^2, \cr
& \simeq\;1\;+\;{1 \over 2}\;W''(0)\;{({\tau \over T})}^2 &\eqnx\qq\cr}$$
and that $W(0)\;=\;1$, $W'(0)\;=\;0$, we get

$$\eqalignno{{\cal F}(\Omega, T)\;& \simeq{\left(1\;-\;{W''(0) \over {2\;
T^2}} {{\partial}^2 \over {{\partial} {\Omega}^2}}\right)}^2\; {\cal F}
(\Omega, \infty)\cr
& \simeq\;{\cal F}(\Omega, \infty)\;-\;{W''(0) \over {T^2}} {{\partial}^2
\over {{\partial} {\Omega}^2}}\left[{\cal F}(\Omega, \infty)\right]&
\eqnx\qq.\cr}$$
This gives the rate

$${\cal R}(\Omega, T)\; =\; {\cal R}(\Omega, \infty)\; -\; {W''(0)
\over {T^2}} {{\partial}^2 \over {{\partial} {\Omega}^2}}\; \left[{\cal
R}(\Omega, \infty)\right]\; +\; O\left({1 \over {T^4}}\right)\eqn\qq$$
for any window function and trajectory.
\medskip
\beginsection{\bf 4. Detector response in Schwarzschild and de-Sitter
coordinate systems}

In this section we shall indicate how the above results can be generalised
to obtain the detector response in Schwarzschild and de-Sitter spacetimes
for observers who are stationed at a constant `radius', in (1+1)
dimensions. The Schwarzschild metric in (1+1)dimensions is

$$ds^2\; =\; (1-{2M \over r})\; dt^2\; - \; {dr^2 \over (1-{2M \over r})}
\eqn\qq$$
and under the transformation $r^*\; =\; r \; + \; 2M \; \ln ({r/ 2M}-1)$
the Schwarzschild metric goes over to the Regge-Wheeler metric

$$ds^2\; =\; (1-{2M \over r}) \; (dt^2-dr^{*2})\eqn\qq.$$
The Kruskal-Szekeres (K-S, hereafter) coordinate system is related to the
Regge-Wheeler(R-W, hereafter) system by the following transformation

$$u\; =\; \exp(r^*/ 4M) \; \cosh({t \over4M});\qquad v\; =\; \exp(r^*/ 4M)
\; \sinh({t \over4M}) \eqn\qq$$
so that the metric in this coordinate system is

$$ds^2\; =\; \left({32M^3 \over r}\right)\; \exp(-r/2M)\; (dv^2-du^2)\eqn
\qq.$$
Thus the metrics in the K-S and the R-W coordinate systems are conformally
flat. Since the lagrangian for the massless scalar field in (1+1)
dimensions is conformally invariant, we can take the mode functions to be
plane waves.
\smallskip
We define a vacuum state with respect to the normal modes of the Kruskal-
Szekeres system and study the response of a detector stationed at a
constant $r^*$ in the tortoise coordinate system. The curves of constant
$r^*$ are hyperbolae in the u-v plane of the K-S coordinates and are
similar to the accelerated trajectories in the Minkowski spacetime. It
turns out that a particle detector stationed at constant $r^*$ responds
in the K-S vacuum in a manner similar to an accelerated detector in the
Minkowski vacuum. This well known result can be obtained as follows. The
Wightman function in the (1+1)dimensional case for plane wave normal modes
is

$$D^+(x,x')\; =\; (-{1 \over 4\pi}) \; \left\lbrace \ln\vert ({(t--t'-i
\epsilon)}^2 - {\vert x-x'\vert}^2)\vert\right\rbrace\eqn\qq$$
which for the case of a constant $r^*$ in the K-S system becomes

$$D^+(x,x')\; =\; ({-1 \over 2\pi}) \; \left\lbrace \ln\vert(2 \; \sinh(
{\Delta t \over 8M} -\; i\epsilon)\vert\right\rbrace\eqn\qq.$$
Since this Green's function is invariant with respect to translations in
the t-coordinate, the transition probability rate for the detector at
constant $r^*$ to get excited to the energy level E from $E_0$ is

$${\cal R}_{K-S}\; =\; \sum_E {\vert {\cal M} \vert}^2 \int_{-\infty}^
\infty d\Delta t\;  e^{-i(E-E_0)\Delta t}\; D^{+}(\Delta t) \eqn\qq$$
where $t$ is the time coordinate in the R-W system. Substituting the K-S
Wightman function for an observer at constant $r^*$ in the above integral
we obtain

$${\cal R}_{K-S}\; =\; -\sum_E\; {\vert {\cal M} \vert}^2  \int_{-\infty}^
\infty d\Delta t\; e^{-i\Omega \Delta t}\; \left\lbrace{1 \over {2\pi}}
\ln (2\; \sinh({\Delta t \over 8M}\; -\; {i\epsilon})\right\rbrace\eqn
\qq.$$
Integrating twice by parts, we get

$${\cal R}_{K-S}\; =\; -\sum_E {\vert {\cal M} \vert}^2  \int_{-\infty}^
\infty d\Delta t\;  e^{-i\Omega \Delta t} {\lbrace{1 \over {2 \pi}}
{(8M\; \Omega\; \sinh({\Delta t \over 8M}\; -\; {i\epsilon})}\rbrace}^
{-2}\eqn\qq$$
which is the familiar integral we have already dealt with. The result
is a Planckian spectrum:

$${\cal R}_{K-S}\; =\; \sum_E {{\vert {\cal M} \vert}^2 \over \Omega}
{1 \over {(e^{8\pi M \Omega} -1)}}\eqn\qq.$$
\smallskip
A similar analysis can be carried out for the case of the de-Sitter
metric

$$ds^2\; =\; (1-H^2r'^2)\; dt'^2\; -\; {dr'^2 \over (1-H^2r'^2)}\eqn
\qq$$
where H is a constant. Defining a new coordinate $r^{'*}$ related to
the de-Sitter $r'$ as

$$r^{'*}\; =\; { 1 \over 2H} \; \ln \left\lbrace \vert{ 1+Hr' \over
1-Hr'} \vert \right\rbrace\eqn\qq$$
we get the metric in this coordinate system $(t',r^{'*})$ to be conformal
to the flat space metric, with

$$ds^2\; =\; (1-H^2r'^2)\; (dt^2 - d{r^{'*}}^2)\eqn\qq.$$
The following transformations

$$u'\; =\; e^{Hr^{'*}}\; \cosh(Ht);\qquad v'\;=\; e^{Hr^{'*}}\;
\sinh(Ht)\eqn\qq$$
when performed yields the flat space metric

$$ds^2\; =\; H^{-2}\; {(1 - Hr)}^2\; (dv'^2\; -\; du'^2)\eqn\qq.$$
\smallskip
Just as constant $r^{*}$ trajectories in K-S sysytem and the accelerated
trajectories in the Minkowski $(x,t)$ plane are hyperbolae, the constant
$r{'*}$ trajectories in the $(u',v')$ are also hyperbolae. The study of
the response of a detector stationed at constant $r^{'*}$ in a vacuum
defined with respect to the normal modes in the $(u', v')$ system is
hence similar to the study of the detector response in Schwarzschild
as done above and we get a Planckian response with a temperature $T={H/
{2\pi}}$.
\smallskip
This analysis can be extended to other trajectories in these spacetimes.
Also the finite time detector response with window functions discussed
in sections {\bf 3(a)}, {\bf 3(b)} and {\bf 3(c)} for the case of
inertial and accelerated frames can be trivially extended to these
two spacetimes for the case of detectors in various trajectories.
\bigskip
\beginsection{\bf 4. Conclusions}

The specific conclusions related to various detector models have been
discussed in the earlier sections wherever appropriate. In this section
we shall touch upon the relevance of the present work in a somewhat broader
context.
\smallskip
In bringing together the principles of quantum theory and general
relativity one notices a major issue of conflict: General relativity is
inherently loval in its description while the conventional formulation
of field theory uses global structures to define even the most primitive
concepts like the vacuum state. This point has been repeatedly made in
the literature related to quantum gravity. However, it should also be
noted that there is another, operational angle to the quantum theory as
well. Quantum mechanics emphasises the role of operational definition of
physical quantities including that of the quantum state. As a matter of
principle the same philosophy should be applicable to the field theory as
well. In other words, one would like to define concepts like vacuum state
etc in field theory using purely operational procedures similar to the
ones used, for example in defining the spin of an electron by using a
magnetic field selector.
\smallskip
It is, however, well known that such procedures are exceedingly difficult
to formulate in the case of a relativistic field. The role of particle
detectors assumes special importance in this context. The work by Unruh
and DeWitt comes closest to the operational definition of quantum states
in field theory. In a simplified sense this detector model captures the
essence of the actual particle detection which takes place in the
laboratory. There is, however, one difficulty in the original Unruh-DeWitt
model. This model uses the definition for particle detection which is
based on asymptotic states. The calculations are done to estimate the
transition probability from past infinity to the future infinity. In any
laboratory context, particle detection is local in both space and time.
\smallskip
The analysis in this paper makes a first attempt in investigating the
possibilty of an inherently local definition of particle detection both
in space and time. We have resolved the difficulties which arise in such a
definition and we have provided general formulas to calculate the response
of detectors which have been coupled to the field only for a finite
interval of time. In a future publication we plan to investigate how these
detectors respond in (3+1) curved spacetime while on geodesic and non-
geodesic trajectories. Since these toy-models mimic the physical situation
as regards locality in space and time, we expect the results to shed some
light on the operational definition of quantum processes in curved spacetime.

\beginsection

\centerline{\bf Appendix A}
The finite time detector response integral for the rectangular window
function is

$${\cal F}(\Omega, T)\; =\; \int_{-2T}^{2T} dx \; e^{-i\Omega x} \; (2T-
\vert x \vert) \; G^+(x)\eqn\qq.$$
For the case of an inertial detector the integrals to be evaluated are

$${\cal F}_{ine1}(\Omega, T)\; =\; ({-2 T \over {4 {\pi}^2}}) \; \int_
{-2T}^{2T} dx \; {e^{-i \Omega x} \over {(x-i \epsilon)}^2}\eqn\qq$$
and

$${\cal F}_{ine2}(\Omega, T)\; =\; ({1 \over {4 \pi^2}}) \; \int_{-2T}^
{2T} dx \; {e^{-i \Omega x} \over {(x-i \epsilon)}^2} \; \vert x \vert
\eqn\qq$$
so that ${\cal F}_{ine}(\Omega, T)\; =\; {\cal F}_{ine1}(\Omega, T)\; +
\; {\cal F}_{ine2}(\Omega, T)$. The integral for ${\cal F}_{ine1}$ can be
evaluated with the aid of a rectangular contour (refer figure 1, ${\cal F}
_{ine1}$) in the lower half of the complex $x$-plane with the vertices
given by $A_{i1}(-2T,0)$, $B_{i1}(2T,0)$, $C_{i1}(2T, -i \infty)$ and
$D_{i1}(-2T, -i \infty)$. Since this contour does not enclose the pole,
by Cauchy's theorem the integral around this closed contour is identically
zero. The value of the integral over the edge  $A_{i1}B_{i1}$ can be
expressed in terms of the integrals over the other edges $B_{i1}C_{i1}$
and $D_{i1}A_{i1}$; the contribution from $C_{i1}D_{i1}$ being zero because
of the vanishing integrand on this edge. Thus

$${\cal F}_{ine1}(\Omega, T)\; =\; ({2 T \over {4 {\pi}^2}}) \left\lbrace
\int_{2T}^{2T-i\infty} dx\; {e^{-i \Omega x} \over {(x-i \epsilon)}^2} \;
+ \; \int_{-2T-i\infty}^{-2T} dx \; {e^{-i \Omega x} \over {(x-i \epsilon)
}^2}\right\rbrace\eqn\qq$$
which after some simple manipulations can be expressed as

$${\cal F}_{ine1}(\Omega, T)\; =\; ({2 T \over {4 {\pi}^2}}) \biggl\lbrace
(i e^{-2i \Omega T}) \int_0^{\infty} dv\; {e^{- \Omega v} \over {(v+
\epsilon +2iT)}^2}\;$$
$$ -\; (i e^{2i \Omega T}) \int_0^{\infty} dv\; {e^{- \Omega v} \over
{(v+ \epsilon -2iT)}^2}\biggl\rbrace\eqn\qq.$$
\smallskip
The term  ${\cal F}_{ine2}$ in the inertial detector response function has
a $\vert x \vert$ term in the integrand and hence has to expressed as a sum
of the integrals over limits $(-2T,0)$ and $(0,2T)$ for evaluation.
Incorporating this result and after some manipulations we obtain

$${\cal F}_{ine2}(\Omega, T)\; =\; ({1 \over {4 {\pi}^2}}) \left\lbrace
\int_0^{2T} dx \; {e^{i \Omega x} x \over {(x + i \epsilon)}^2} \; + \;
\int_0^{2T} dx \; {e^{-i \Omega x} x \over {(x - i \epsilon)}^2} \right
\rbrace\eqn\qq.$$
The first of these integrals can be evaluated on a rectangular contour
(refer figure 2, ${\cal F}_{ine2A}$
) in the upper-half of the complex $x$-plane with the vertices at $A_{i2}(0,
0)$, $B_{i2}(2T, 0)$, $C_{i2}(2T, i\infty)$ and $D_{i2}(0, i\infty)$.
Similarly the second integral can be performed with the aid of another
rectangular contour(refer figure 3, ${\cal F}_{ine2B}$), this time in the
lower-half of the complex $x$-plane with the vertices at  $A_{i2*}(0, 0)$,
$B_{i2*}(2T, 0)$, $C_{i2*}(2T, -i\infty)$ and $D_{i2*}(0, -i\infty)$. Since
neither of these contours enclose any poles the integral of consequence
can be expressed in terms of integrals over the edges $B_{i2/2*}C_{i2/2*}$
and $D_{i2/2*}A_{i2/2*}$ alone, the integrand vanishing on the edge
$C_{i2/2*}D_{i2/2*}$. After some simple algebra we obtain

$${\cal F}_{ine2}(\Omega, T)\; =\; ({1 \over {4 {\pi}^2}}) \biggl\lbrace
(2iT \; e^{2i \Omega T})\int_0^{\infty} dv\; {e^{-\Omega v} \over {(v+
\epsilon -2iT)}^2} $$
$$ - \; (2iT \; e^{-2i \Omega T}) \int_0^{\infty} dv \; {e^{-\Omega v}
\over {(v+ \epsilon +2iT)}^2} \; - \; e^{2i \Omega T} \int_0^{\infty}
dv \; {e^{-\Omega v} v \over {(v+ \epsilon -2iT)}^2} \;$$
$$ - \; e^{-2i \Omega T} \int_0^{\infty} dv \; {e^{-\Omega v} v \over
{(v+ \epsilon +2iT)}^2} \; + \; 2 \int_0^{\infty} dv \; {e^{-\Omega v}
v \over {(v+ \epsilon)}^2}\biggl\rbrace \eqn\qq$$
so that the finite time inertial detector response is
$${\cal F}_{ine}(\Omega, T)\; =\; {1 \over {4 {\pi}^2}} \biggl\lbrace -
e^{2i \Omega T} \int_0^{\infty}dv\;  {e^{-\Omega v} v \over {(v+ \epsilon
-2iT)}^2} - e^{-2i \Omega T} \int_0^{\infty} dv\; {e^{-\Omega v} v \over
{(v+ \epsilon +2iT)}^2}$$
$$+ 2 \int_0^{\infty}dv\; {e^{-\Omega v} v \over {(v+ \epsilon)}^2}\biggl
\rbrace\eqn\qdetressit.$$
This is the result quoted earlier in the text.
\bigskip
\beginsection

{\centerline{\bf Appendix B}
For the case of an accelerated trajectory the finite time detector integral
with the rectangular window function is

$${\cal F}_{acc}(\Omega, T)\; =\; \sum_{n=-\infty}^{\infty} \left\lbrace
{\cal F}_{acc1n}(\Omega, T) + {\cal F}_{acc2n}(\Omega, T)\right\rbrace
\eqn\qq$$
where

$${\cal F}_{acc1n}(\Omega, T)\; =\; {-2T \over {4 \pi^2}} \int_{-2T}^{2T}
dx \; {e^{-i \Omega x} \over {(x-ib_n)}^2}\eqn\qq$$
and

$${\cal F}_{acc2n}(\Omega, T)\; =\;{1 \over {4 \pi^2}} \int_{-2T}^{2T} dx
\; {e^{-i \Omega x} \; \vert x \vert \over {(x-ib_n)}^2}\eqn\qq.$$
${\cal F}_{acc1n}(\Omega, T)$ can be evaluated with the aid of a rectangular
contour (refer figure 4, ${\cal F}_{acc1}$) with the vertices at $A_{a1}
(-2T,0)$, $B_{a1}(2T,0)$, $C_{a1}(2T, -i \infty)$ and $D_{a1}(-2T, -i
\infty)$. This contour encloses the poles corresponding to the values of
n between one and infinity and the integral for
${\cal F}_{acc1n}(\Omega, T)$ can be expressed in terms of the integrals
over the edges $B_{a1}C_{a1}$ and $D_{a1}A_{a1}$ and the residues
corresponding to the enclosed poles. After some manipulations we obtain

$${\cal F}_{acc1n}(\Omega, T)={2T \over {4 \pi^2}} \biggl\lbrace 2 \pi
\Omega \Theta(n) e^{\Omega b_n} + (ie^{-2i\Omega T}) \int_0^{\infty} dv
{e^{-\Omega v} \over {(v+b_n +2iT)}^2}$$
$$ -(ie^{2i\Omega T}) \int_0^{\infty} dv {e^{-\Omega v} \over {(v+b_n -
2iT)}^2}\biggl\rbrace \eqn\qq$$
where $\Theta(n)\;=\;1$ for $n\;>\;0$ and zero otherwise.
\smallskip
${\cal F}_{acc2n}(\Omega, T)$, after having been split into two integrals
with the limits $(-2T,0)$ and $(0,2T)$ reduces to

$${\cal F}_{acc2n}(\Omega, T)= {1 \over {4 \pi^2}} \left\lbrace \int_0^
{2T} dx {e^{i\Omega x} x \over {(x+ib_n)}^2} + \int_0^{2T} dx {e^{-i\Omega
x} x \over {(x-ib_n)}^2}\right\rbrace\eqn\qdetresatwosfn.$$
The first of these integrals can be performed with the help of a rectangular
contour (refer figure 5, ${\cal F}_{acc2nA}$) on upper-half of the complex
$x$-plane with the vertices $A_{a2}(0, 0)$, $B_{a2}(2T,0)$, $C_{a2}(2T, i
\infty)$ and $D_{a2}(0, i \infty)$; but for the cases when $n>0$ the pole
in the integrand sits right on the edge $D_{a2}A_{a2}$ and to avoid it we
indent the contour in such a way so that the pole is left outside.
Similarly for evaluating the second integral in \qdetresatwosfn\ a contour
(refer figure 6, ${\cal F}_{acc2nB}$) with vertices $A_{a2*}(0, 0)$,
$B_{a2*}(2T,0)$, $C_{a2*}(2T, -i \infty)$ and $D_{a2*}(0, -i \infty)$ can
be chosen and the poles sprouting on the edge $D_{a2*}A_{a2*}$ for the
values of n between one and infinity can be avoided with an indentation
as to leave them outside. The indentation on the contours contribute a
residue corresponding to the infinitesimal semicircle around the pole and
only the principal value of the integral over the edges $D_{a2}A_{a2}$ and
$D_{a2*}A_{a2*}$ can be defined with the result

$${\cal F}_{acc2n}(\Omega, T)={1 \over {4 \pi^2}} \biggl\lbrace (2iT e^{2i
\Omega T}) \int_0^{\infty} dv {e^{-\Omega v} \over {(v+b_n -2iT)}^2}$$
$$ - e^{2i \Omega T} \int_0^{\infty} dv {e^{-\Omega v} v \over {(v+b_n -
2iT)}^2}-(2iT e^{-2i \Omega T}) \int_0^{\infty} dv {e^{-\Omega v} \over
{(v+b_n +2iT)}^2}$$
$$-e^{-2i \Omega T} \int_0^{\infty} dv {e^{-\Omega v} v \over {(v+b_n -
2iT)}^2}+ 2 \int_0^{\infty} dv {e^{-\Omega v} v \over {(v+b_n)}^2}\biggl
\rbrace\eqn\qq.$$
It is assumed that when the pole happens to settle right on the axis of
integration the integral over the axis is taken to be its principal
value. The complete accelerated detector response is then given by

$${\cal F}_{acc}(\Omega, T)= {1 \over 4\pi^2} \sum_{n=-\infty}^{\infty}
\biggl\lbrace 2 \pi 2T \Omega \Theta(n) e^{\Omega b_n} -  e^{2i \Omega
T} \int_0^{\infty} dv {e^{-\Omega v} v \over {(v+b_n -2iT)}^2}$$
$$ - e^{-2i \Omega T} \int_0^{\infty} dv {e^{-\Omega v} v \over {(v+b_n
+2iT)}^2} + 2 \int_0^{\infty} dv {e^{-\Omega v} v \over {(v+b_n)}^2}
\biggl\rbrace\eqn\qdetressat$$
as quoted earlier.
\vfill\eject
{\bf References}

[1] S. A. Fulling, Phys. Rev. D {\bf 10}, 2850 (1973).
\medskip
[2] W. G. Unruh, Phys. Rev. D {\bf 14}, 870 (1976).
\medskip
[3] B. S. DeWitt, in {\sl General Relativity: An Einstein Centenary
Survey}, edited by S. W. Hawking and W. Israel (Cambridge University
Press, Cambridge, England, 1980).
\medskip
[4] P. C. W. Davies, in {\sl Quantum theory of Gravity}, edited by S.
M. Christensen (Hilger, Bristol, 1984).
\medskip
[5] B. F. Svaiter and N. F. Svaiter, Phys. Rev. D {\bf 46}, 5267
(1992).
\medskip
[6] A. Higuchi, G. E. A. Matsas, C. B. Peres, Phys. Rev. D {\bf 48},
3731 (1993).
\medskip
[7] I. S. Gradshetyn and I. M. Ryzhik, {\sl Table of Integrals, Series
and Products} (Academic, New York, 1980).
\medskip
[8] T. Padmanabhan, Astrophys. Space. Sci. {\bf 83}, 247 (1982).
\medskip
[9] N. D. Birrell and P. C. W. Davies, {\sl Quantum Fields in Curved
Space}  (Cambridge University Press, Cambridge, England, 1982).
\end